\documentclass{aastex61}

\include{amsmath}

\received{May 24, 2018}
\revised{May 24, 2018}
\accepted{\today}
\submitjournal{ApJ}

\shorttitle{Ionizing Radiation Field}
\shortauthors{Henry et al.}

\begin{document}

\title{Discovery of An ionizing Radiation Field in the universe}

\correspondingauthor{Richard Conn Henry}
\email{henry@jhu.edu}

\author[0000-0002-0786-7307]{Richard Conn Henry}
\affiliation{The Johns Hopkins University, 
3400 North Charles Street, 
Baltimore, MD 21218-2686, USA}

\author{Jayant Murthy}
\affiliation{Indian Institute of Astrophysics, Bengaluru, India}
%\nocollaboration

\author{James Overduin}
\affiliation{Towson University, Baltimore, MD, USA}
%\nocollaboration

\begin{abstract} % Abstract

We draw attention to observational evidence indicating that a substantial fraction of the well-known cosmic celestial diffuse ultraviolet background radiation field is actually due not to dust-scattered starlight, but rather---considering its spectral character at most locations in the sky---has an unknown physical origin.  We arrive at this conclusion from re-examination of spectra of the diffuse ultraviolet background that were obtained---long ago---using the ultraviolet spectrometers aboard the two Voyager spacecraft, which were located far out in our solar system, and at very different locations.  As there is neither a reasonable nor even an unreasonable conventional astrophysical source for this newly-identified fraction of the radiation, we are led to speculate that the photons that we observe have their origin in the very slow decay of the particles that make up the ubiquitous dark matter, which we know envelopes our Galaxy.  Whether or not that actually is the source, this new radiation field extends somewhat below 912 {\AA}, so we have found, at last, the radiation that re-ionized the universe.

\end{abstract}


\section{Introduction} \label{sec:intro} % Introduction

William of Ockham (c. 1287-1347) was, according to Wickipedia, ``an English Franciscan friar, scholastic philosopher, and theologian" who---and wisely---advocated tentative acceptance of the simplest possible explanation for observed phenomena.
The astrophysical community has largely followed that sage advice, in particular believing the diffuse cosmic ultraviolet background to originate overwhelmingly in the scattered light of OB stars in our Galactic neighborhood (Hamden et al. 2013).
This position has been easy to adopt, as there has been no known alternative source for such radiation that could be bright enough to explain the observed background.  The discovery (Meekins et al. 1971; Gursky et al. 1971) that the dark matter is non-baryonic, however, drastically changes the situation:  our galaxy is immersed in dark matter that conceivably could slowly decay with the emission of photons in the spectral range 900\,\AA\  to 1400 \AA\ (or so) for which the Voyager missions' ultraviolet spectrometers were highly sensitive (Ben-Jaffel and Holberg 2016) but for which the GALEX mission's far ultraviolet photometry (1330 {\AA} - 1750 {\AA})  was largely (but not completely) insensitive (Hamden et al. 2013), and this might provide a background as bright as what is observed.

Peculiar Voyager observations have been reported by Murthy et al. (1999, 2012).  Their utterly mysterious character has been emphasized by Henry (2003).  We present these observations once again in the present paper because of their, we believe, extreme importance, both for astronomy, and for fundamental physics.  

We furthermore propose a simple and definitive test of these photons' physical origin through observations using the Alice ultraviolet spectrometer (Gladstone, Stern, and Pryor 2013) aboard the New Horizons spacecraft now beyond Pluto, to (we hope and expect), with Alice's 10 \AA\ spectral resolution, rule out any possibility that the radiation is dust-scattered starlight, or instrumental scattering of the light of off-axis bright OB stars.

\begin{figure}[ht!] % FIGURE ONE
\plotone{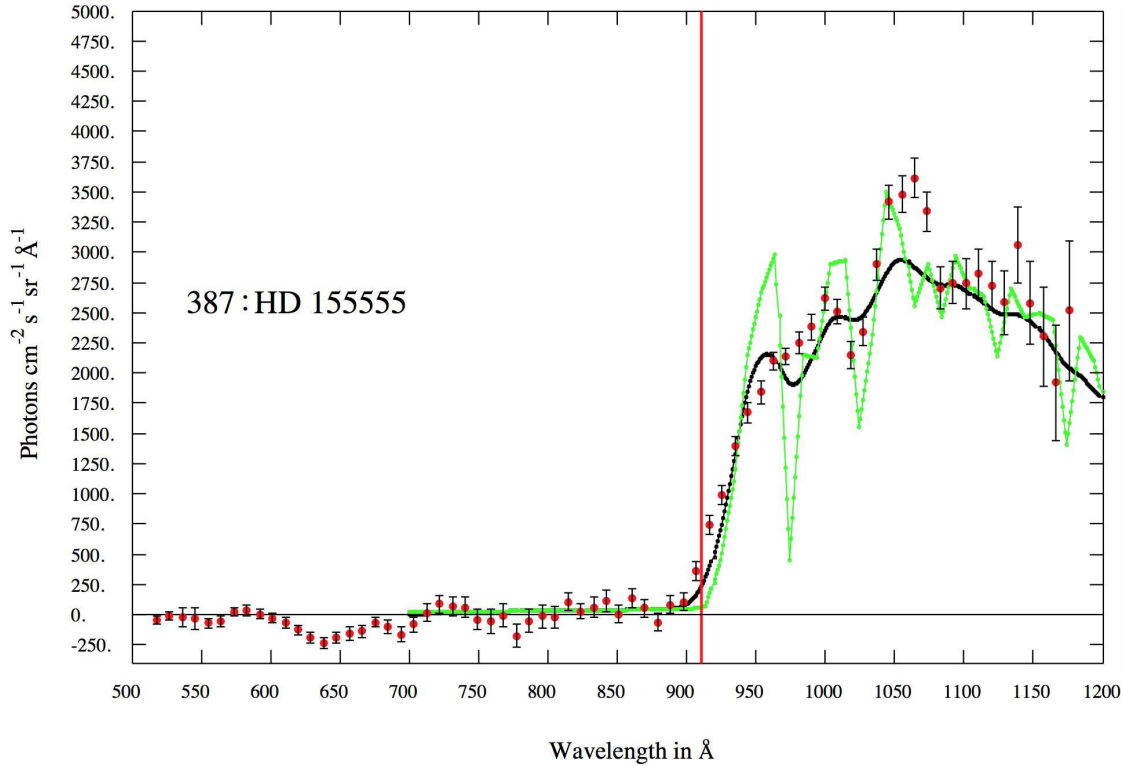}
\caption{The spectrum of the diffuse ultraviolet background, as observed with the spectrometer aboard Voyager 2, reproduced from the web page http://henry.pha.jhu.edu/voyager.html  ; this is spectrum  \#387 in that collection ($\ell$ = 324.9, {\it b} = -16.3).  The label HD 155555 is entirely meaningless---being only a name.  The observation is of a blank region of the sky, and as the signal was found not to vary significantly as the pointing varied, the signal  must be due to diffuse radiation, with little contribution from any point sources.  A comparison spectrum of a model of an early-type star (25,000$^{\circ}$) is also shown, both at the model's full spectral resolution, and also as degraded to the 38  {\AA} resolution of the Voyager spectrometer.  The excellent fit convinces us that for this particular target---unlike for most of the Voyager background spectra---we are indeed observing mostly dust-scattered starlight.  The Alice spectrometer aboard New Horizons has 10  {\AA} resolution, which would guarantee identification of cosmic background spectra that are dominated by the scattered light of early-type stars. }
\end{figure}

\section{History} \label{sec:intro} % History

Observation of the cosmic ultraviolet background began with Henry (1973) and references therein.  The subject has been reviewed by Bowyer (1991) and by Henry (1991), who came to very different conclusions:  Bowyer's view was that the bulk of 
the diffuse FUV is starlight scattered from dust, while Henry claimed that the bulk of the radiation is extragalactic.  

There is no question that hot stars in our neighborhood in the Galaxy produce copious amounts of EUV and FUV radiation,
nor that there is plenty of interstellar dust that might well be scattering that radiation into our rocket-borne and satellite-borne detectors.  The only question is, how much?  That depends entirely on the values of the interstellar dust-grain albedo $\it {a}$ and their Henyey-Greenstein (1941) scattering 
parameter $\it {g}$:  which values one attempts to determine, by observing the celestial distribution of the observed FUV background, and fitting models---the models test whether the scattering hypothesis can
account for the observed distribution of diffuse FUV radiation on the sky.  If it can, Ockham advocates that that conclusion be tentatively adopted, and quite properly, in recent years, it has.  

\begin{figure}[ht!] % FIGURE TWO
\plotone{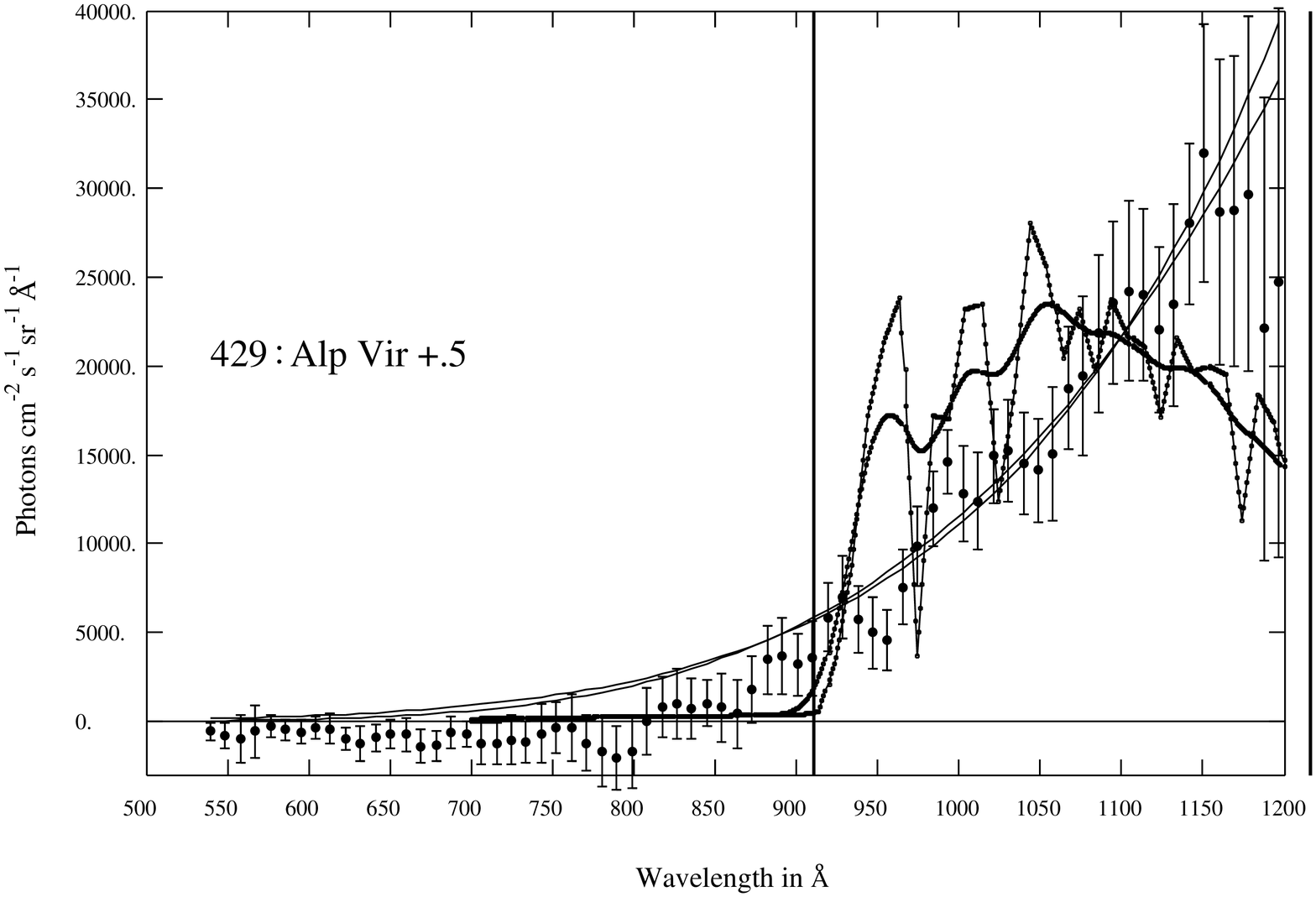}
\caption{The Voyager 1 spectrum of \#429 (Table 1) is compared with the same early-type stellar spectrum that appeared in Figure 1.  The fit to the data is terrible.  We reject the notion that the diffuse background
observed at this location in the sky is starlight scattered from dust.  The great majority of the Voyager spectra (Voyager 1 and Voyager 2) are of similar to what appears in this figure.  This means that
the great bulk of the diffuse ultraviolet background radiation far from the Galactic equator is due not to starlight scattered from dust, as has been widely believed, but rather has a totally unknown origin.  
The  spectrum is seen to drastically increase   (   from perhaps 2000 photons cm$^{-2}$ s$^{-1}$ sr$^{-1}$ \AA $^{-1}$ at 912 \AA\   )  toward longer wavelengths, and is still rising as 1216 {\AA}  Lyman $\alpha$ is approached.  
What happens longward of Lyman $\alpha$ is shown in Figure 3.}
\end{figure}

\section{The GALEX Diffuse Ultraviolet Cosmic Background} \label{sec:intro} % The GALEX Diffuse Ultraviolet Cosmic Background

Fortunately, the cosmic diffuse ultraviolet background radiation status has recently been greatly clarified by the success of the GALEX mission.  Hamden et al. (2013) found that the ultraviolet background could be fit over the far ultraviolet with a model for the dust-scattered light of OB stars in our cosmic neighborhood---but they found that an additional component of unknown origin is needed at high galactic latitudes.  Their data were reexamined by Henry et al. (2015) who agreed with their conclusion, but who emphasized the importance of the fact that the highest-Galactic latitude observations did not fit the model, so that, as the signal was now demonstrated to be 2-component, the dust-scattered starlight component could no longer be used to set the values of the albedo {\it a} and Henyey-Greenstein scattering parameter {\it g}.  We have recently critically reassessed (and confirmed) our determination of the requirement for an unknown second component for the GALEX FUV background (Akshaya et al. 2018).  The question now becomes, what fraction of the total diffuse ultraviolet background comes from dust-scattered starlight, and what fraction comes from the unknown second source?  We will be pointing out below the practical possibility of making simple and definitive future observations that can easily distinguish between dust-scattered starlight and any other sources for the UV background, because the dust-scattered starlight component necessarily has strong stellar-atmosphere absorption lines.

\begin{figure}[ht!]  % FIGURE THREE
\plotone{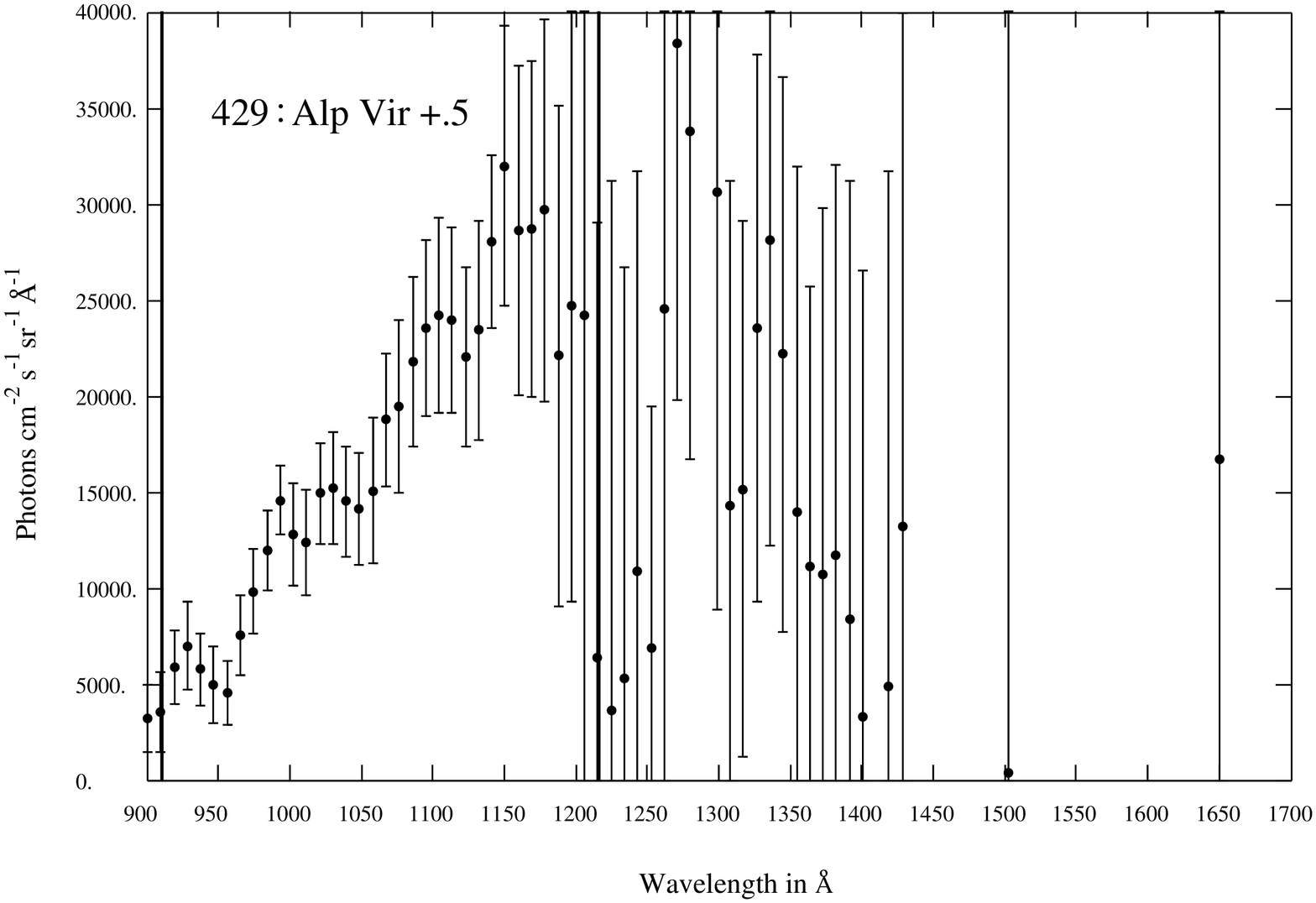}
\caption{The remarkable spectrum that appeared in Figure 2 does {\it not} continue to increase beyond Lyman $\alpha$.  Unfortunately the Voyager spectrometers had little sensitivity beyond Lyman $\alpha$, 
but---and very fortunately indeed---they did have just enough sensitivity to allow us to conclude that the peculiar ultraviolet background spectra that we find shortward of Lyman $\alpha$ do not continue longward of (about) Lyman $\alpha$---which explains why they do not appear in the GALEX images  of the diffuse ultraviolet background radiation presented by Hamden et al. (2013).  Notice the flux scale: the brightest general diffuse background reported by Henry et. al. (2015) was, with the lowest Galactic latitudes avoided,  8962 photons cm$^{-2}$ s$^{-1}$ sr$^{-1}$ \AA $^{-1}$.}
\end{figure}

\begin{figure}[ht!]   % FIGURE FOUR
\plotone{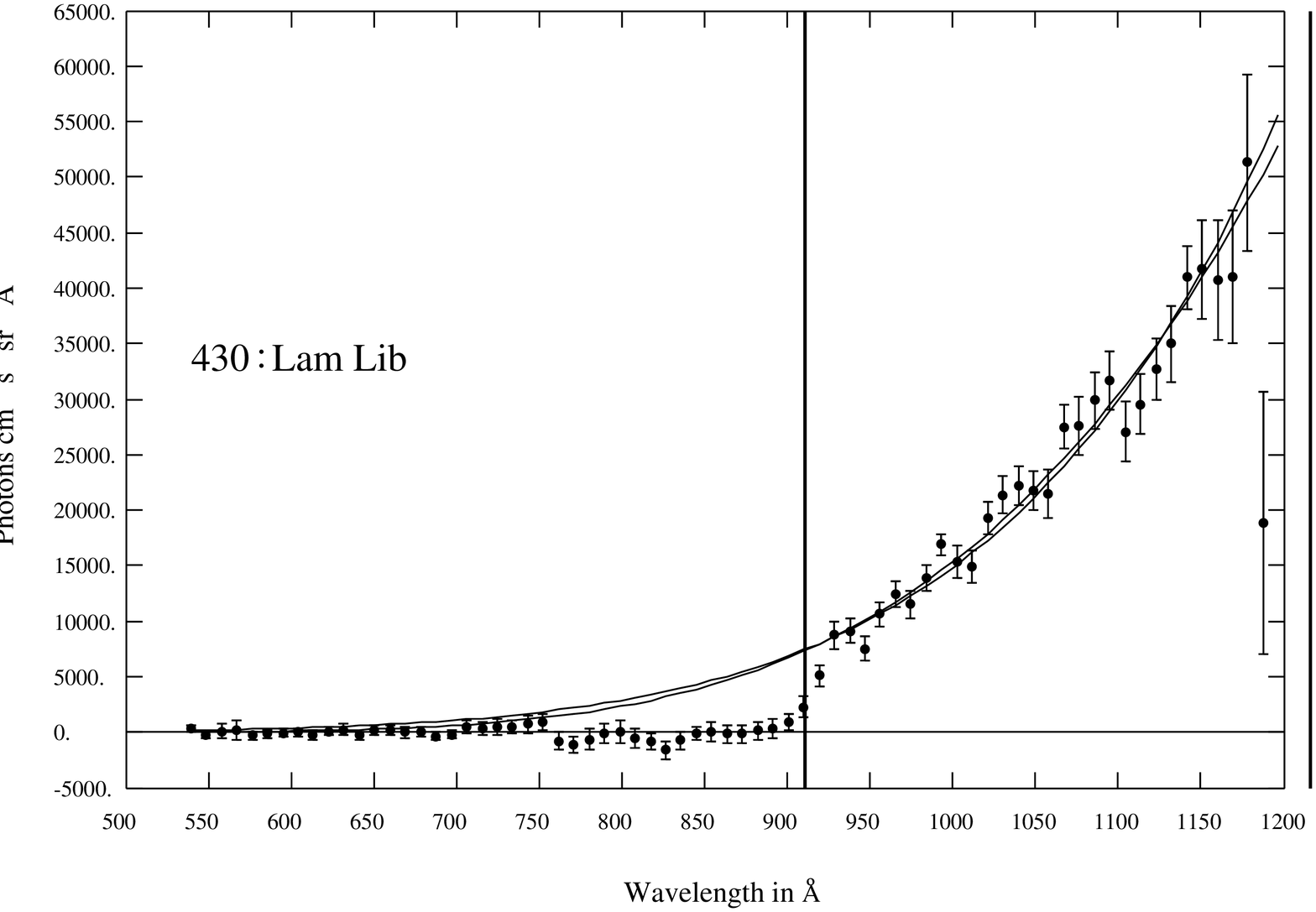}
\caption{A second example (\#430) of the truly remarkable Voyager spectra of the diffuse ultraviolet background radiation shortward of (about) Lyman $\alpha$.   The brightest general diffuse background reported by Henry et al. (2015) was 8962 units.  In Figure 2 a trace of structure appeared in the sharply rising spectrum, suggesting that there might be some contribution of OB starlight (which has strong absorption lines in its spectrum) scattered from interstellar dust.  But for \#430 the spectrum is smooth and is well-fitted by power-law or exponential spectra (shown).  The spectrum is cut off at the 912 {\AA} Lyman limit because of the interstellar hydrogen that lies between us and the dark-matter source.  But it is hardly conceivable that that source does not continue to shorter wavelengths, easily providing the hard radiation that re-ionized the universe following recombination.  }
\end{figure}

\section{voyager spectra} \label{sec:voyager spectra}   %  voyager spectra
In Table 1 we list the 430 Voyager observations, some of which we will be displaying in order to bring out their remarkable properties.  The first column in the table is a serial number; the second column, the number in Murthy et. al. (1999).  Subsequent columns give the celestial and Galactic coordinates of each observation, and identify which of the two Voyager spectrometers made that particular observation; a number characterizing the UV brightness observed (on a linear scale); and finally a nickname assigned by the original observers (most of the observations were their off-target reference measurements). 

What was expected to be seen  in these spectra (in the dust-scattering interpretation for the origin of the ultraviolet background)  was of course the light of OB stars in our Galactic neighborhood scattered off of  interstellar dust, and in Figure 1 we demonstrated that the Voyager spectrometers were quite capable of delivering the goods, and that they did so:  the spectrum is clearly that of an early-type star (and we also have learned from this observation that the  albedo {\it a} and Henyey-Greenstein scattering parameter {\it g} do not vary significantly with wavelength).

However, the vast majority of the spectra reported in Table 1 do {\it not} have this (the expected) character---in fact, they are radically different.  In Figure 2 we show the spectrum of target \#429 from Table 1, and we include in the figure the same OB star spectral shape that appears in Figure 1.  It is quite clear that we are now not seeing dust-scattered starlight, but something radically different.  The figure includes power-law and exponential spectra which do fit the observation.  That indicates a different, and entirely unexpected, origin for this radiation---an origin, we suggest, in the slow decay of dark matter particles---that apparently being the sole alternative possibility.  And we think that William of Ockham would approve.

The observed spectrum shown in Figure 2 increases sharply toward longer wavelengths.  Why, then, was this source (and many similar sources) not seen in the images of Hamden et al. (2013)?  Fortunately, the Voyager spectrometers had just enough sensitivity at wavelengths longward of Lyman $\alpha$ to allow us to answer that question:  in Figure 3, we show the entire \#429 spectrum, and we see that the brightness falls abruptly and dramatically longward of about 1200\,{\AA}, calling for a truly remarkable physical source for the radiation---thus once again suggesting the probability that we are seeing dark matter decay.

  \section{the ionization of the universe} \label{sec:ionization}  % the ionization of the universe
  
Following its well-established recombination (after 379,000 years of rapid expansion) we have no doubt at all that the universe managed, somehow, to get itself almost entirely reionized.  What was the source of the necessary ionizing photons?  A good question---for ionizing almost all of the baryonic matter in the universe is not a trivial undertaking, as pointed out by Kollmeier et al. (2014) in their paper, ``The Photon Underproduction Crisis."  Subsequent to the appearance of that paper there has been much activity to ``fix" the situation by exploring ways that ionizing photons could escape from small galaxies in the early universe;  also,  Borthakur et al. (2014) have found that in some cases ionizing photons do actually escape from even late-forming galaxies.  But the question of how almost all the baryonic matter in the universe got reionized is, today, very far from being settled.  

However, if indeed the sources described in the present paper emit any significant fraction of their radiation below the Lyman limit of 912\, {\AA}, and the spectrum of Figure 4 indicates that that is so, the question becomes moot:  we have discovered how the universe became reionized.  Figure 4 shows a spectrum that clearly continues shortward of the 912\,{\AA} Lyman limit.  At our location near the Galactic plane, we are immersed in neutral hydrogen gas, so we cannot see such radiation, but this spectrum clearly implies that it exists.  Being created well above the galactic plane, the radiation is available to ionize any neutral hydrogen that is outside the disk of our Galaxy.  And of course the slowly-decaying dark matter is omnipresent in the Universe.

  \section{Conclusion} \label{sec:conclusion} % Conclusion
  
 We propose that the universe was reionized (following its recombination) by the decay of dark matter particles emitting a continuum of photons in the range  $\sim$ 850  {\AA} to about 2000 {\AA}, with the bulk of the radiation being emitted over the range 1000  {\AA} to  1200 {\AA} (a range not observed by GALEX).  We have detected the emission (using the Voyager ultraviolet spectrometers), in various directions in our neighborhood of the Galaxy, with (for example) a striking concentration in the direction just north of the Galactic center, as appears in Figure 6, and an even more probative example---because it is much farther from Gould's belt---the emission we see at  $\ell$ = 190$^{\circ}$, {\it b} = +30$^{\circ}$ to +50$^{\circ}$.
 
In the first figure of this paper we showed that the Voyager spectrometers {\it can} return excellent measurements of what is clearly and unequivocally dust-scattered starlight.  An important second example of such observations is given in Murthy et al. (1994), being observations of the Coalsack nebula---which is interstellar dust in great quantities illuminated by by some of the brightest UV-emitting stars in the sky:  $\alpha$ Crucis (B0.5 IV + B1 V; V = 0.76), $\beta$\,Centauri (B1 III + B1 II + B1 V?; V = 0.61), $\beta$ Crucis (B0.5 IV / B1 V; V = 1.25), and $\delta$ Crucis (B2 IV; V = 2.79).  There, the Voyager spectrometers (both of them) easily and clearly see just what was generally expected to be seen all over the sky---but what is emphatically {\it not} seen in the great majority of the Voyager spectra.  And there is no trace of dark matter emission, despite the low Galactic latitude of these observations.

We have, we hope understandably, been quick to publish what confirmed general expectations, but rather slower to publish what confounds those expectations.  In this paper, we are making, of course,  extraordinary claims, and we are acutely aware that Carl Sagan warned ``that extraordinary claims require extraordinary evidence."  We emphatically agree with Sagan and the principal purpose of this paper is to  draw attention to the fact that conclusive evidence would be easy to obtain if we could make simple observations with the Alice ultraviolet spectrometer aboard the New Horizons mission, which is now well past the planet Pluto and is therefore relatively free from the grating-scattered solar Lyman $\alpha$ that would otherwise be a serious impediment to such measurements.  We hope very much that such measurements will be made shortly after the 2019 January 1 New Horizons investigation of Kuiper belt object 2014 MU69.  A few long integrations with Alice at the Galactic poles and some lower Galactic latitude locations would confirm---or not---the discovery that we are convinced that we have made.
 
 Finally, Henry, Murthy, Overduin, and Tyler (2015) made an effort to find within and beyond standard model physics some dark matter source that could possibly provide the observed GALEX ultraviolet radiation.  We had no success in that good-faith effort.  How then is one to truly identify the physical source of the radiation we report in this paper?  Compare with the sun:  how do we convince ourselves that daylight comes from the sun?  The light comes from the same direction from which comes the gravitational attraction that moves the Earth.  That convinces us!  In the same way, we hope that in the future the detailed location of the dark matter (that we are certain is the dominant feature of our Galaxy) can be measured {\it gravitationally} (with the help of Gaia) with sufficient accuracy to correlate it with the locations we have identifed where strong diffuse ultraviolet emission is somehow (new physics) being generated.  Until that day our conclusion must remain a guess---but we think it is a good one.
 
 \begin{figure}[ht!]   % FIGURE FIVE
\plotone{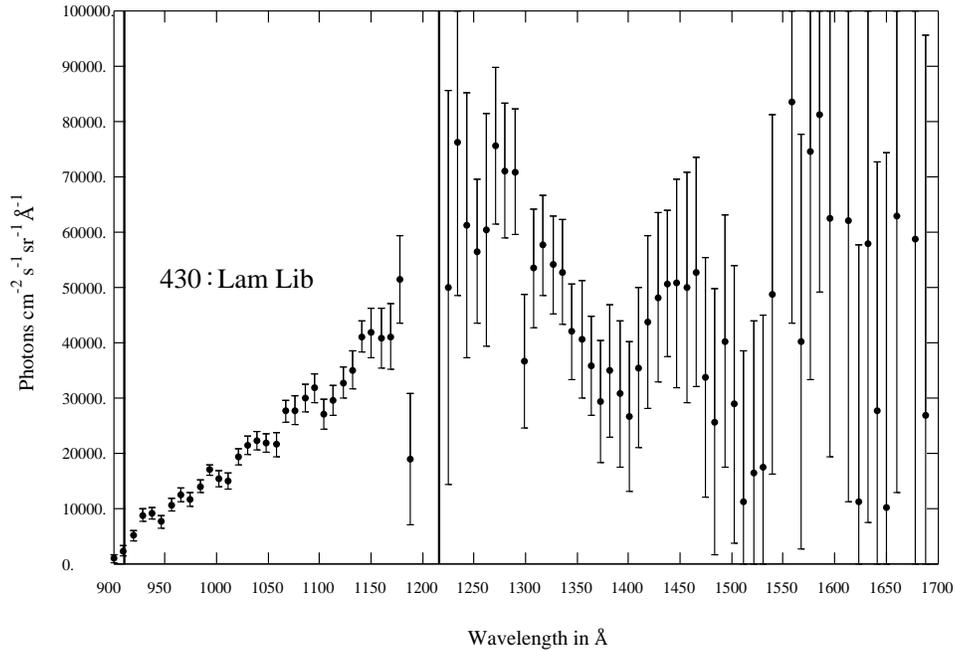}
\caption{Just as with spectrum \#429, spectrum \#430 does not continue its rapid rise much beyond Lyman $\alpha$. This unusual spectrum should be a powerful diagnostic as to the new fundamental physics involved in its emission. }
\end{figure}
%\clearpage

\begin{figure}[ht!]   % FIGURE SIX
\plotone{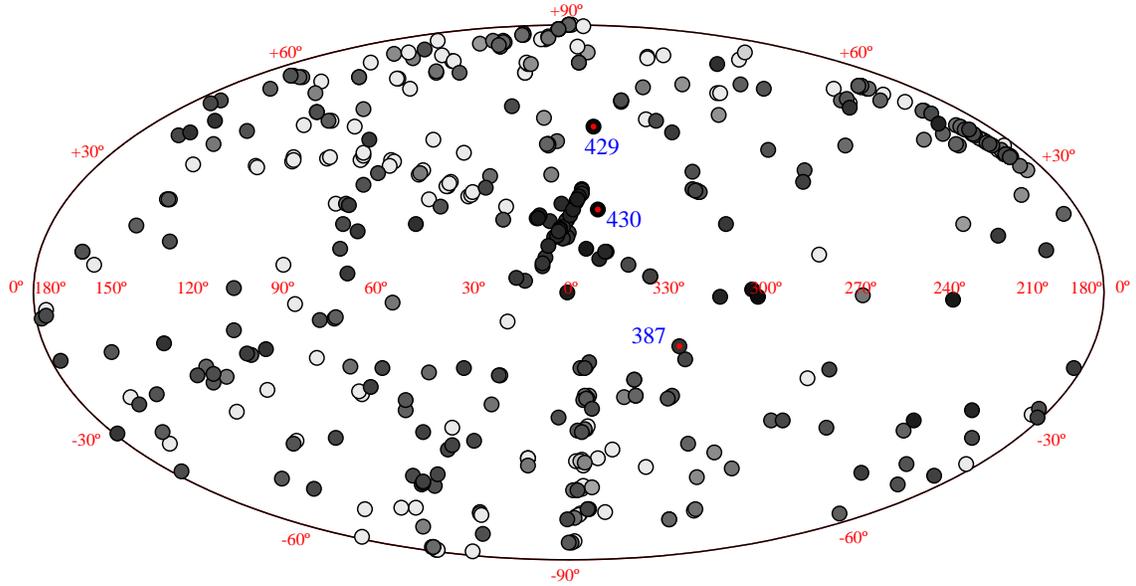}
\caption{The 430 Voyager UV brightnesses ($\sim$ 950   {\AA}  - 1150  {\AA})  range from almost zero to $\sim$34,000.  The symbols are shaded from faintest (very light gray) to brightest (black), with the grayscale being logarithmic.  We see a dramatic concentration of dark matter in a direction just above the Galactic center. An almost equally impressive concentration (and one that is also seen by both Voyager\,1 and Voyager\,2) appears at $\ell$ = 190$^{\circ}$, {\it b} = +40$^{\circ}$.  Radiation at these short wavelengths is strongly absorbed by the interstellar medium, so we do not see more than at most a few hundred parsecs at low Galactic latitudes.  (The Voyager spectrometers' fields of view are $0\fdg1 \times 0\fdg87  $, displayed as circles for visibility.)}
\end{figure}

\begin{figure}  % FIGURES SEVEN AND EIGHT
    \centering
    \begin{minipage}{0.45\textwidth}
        \centering
        \includegraphics[width=0.9\textwidth]{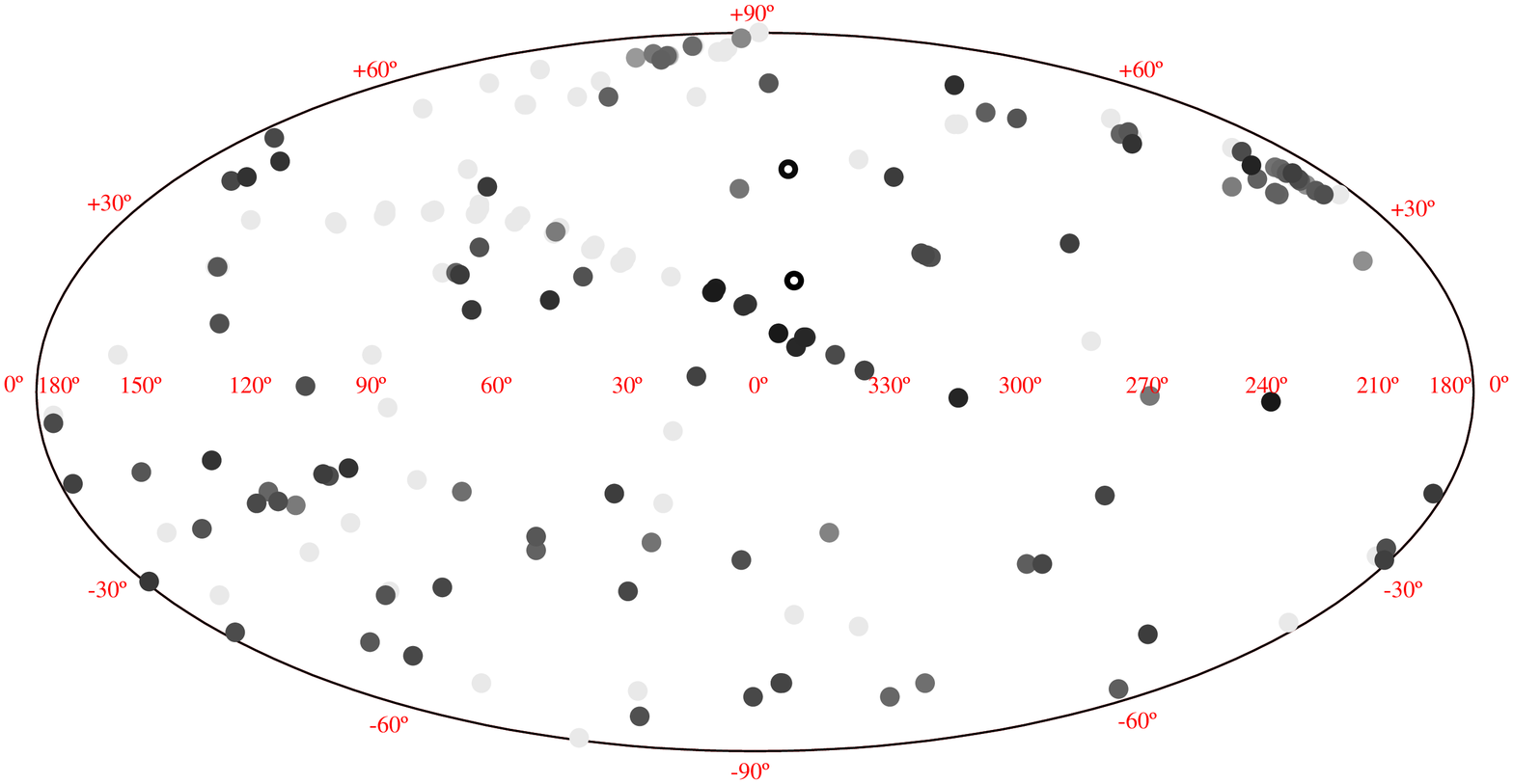} % first figure itself
        \caption{The 186 Voyager \,1 UV brightnesses from our Figure 6 are shown here separately in order to test for consistency between what is observed with  Voyager \,1 and what is observed, from a very different location in the solar system, with the Voyager \,2 ultraviolet spectrometer.}
    \end{minipage}\hfill
    \begin{minipage}{0.45\textwidth}
        \centering
        \includegraphics[width=0.9\textwidth]{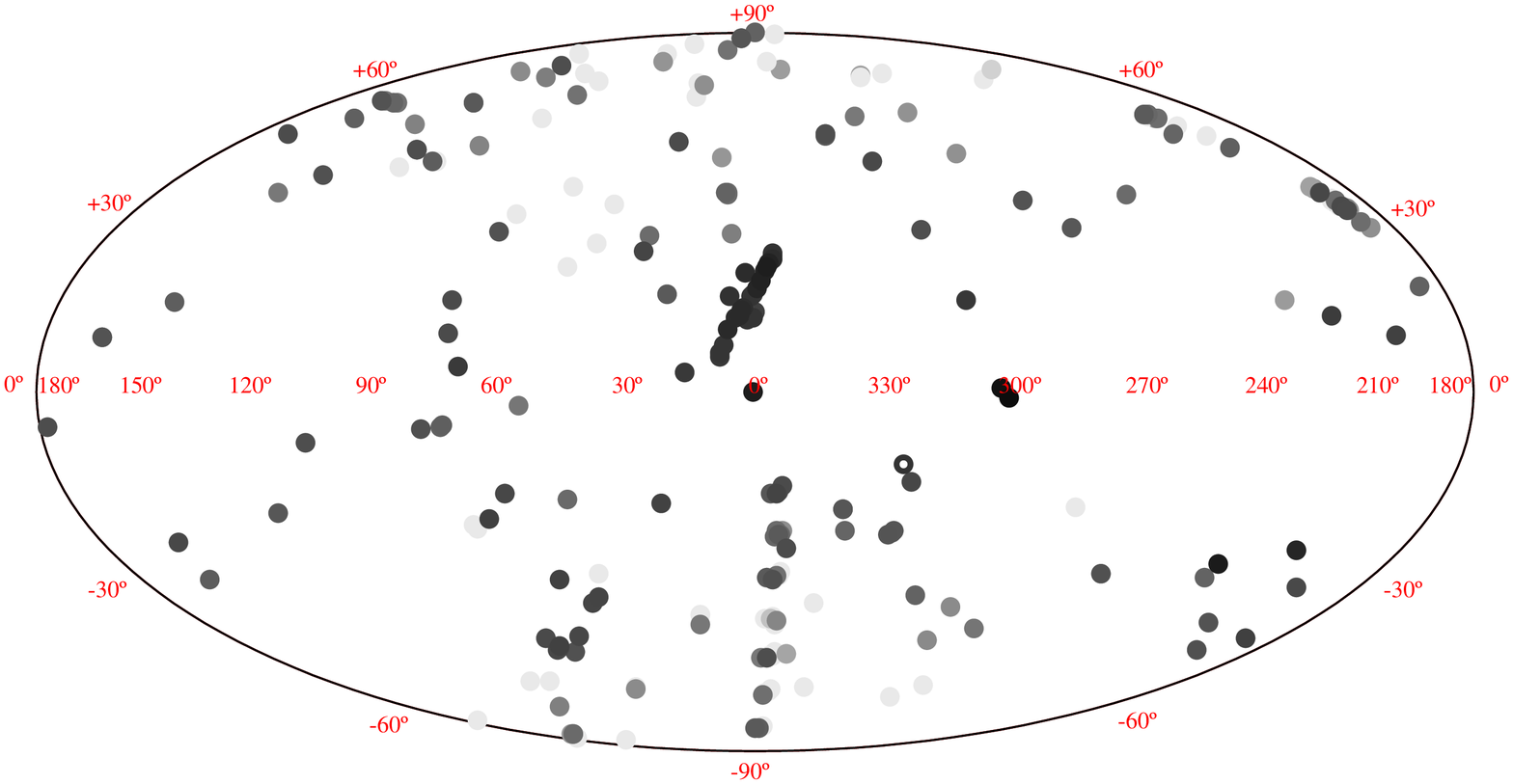} % second figure itself
        \caption{The 244 Voyager \,2 UV brightnesses from our Figure 6.  There is clearly general agreement with the Voyager \,1 observations of Figure 7, giving confidence that both are reporting real signals---images of the dark matter locations, as measured from our position in the galaxy.}
    \end{minipage}
\end{figure}

 \begin{deluxetable*}{ccccccccl}[b!]   % our REAL TABLE 
 \tablecaption{The 430 Voyager Spectra}
 \tablecolumns{9}
 \tablenum{1}
  \tablewidth{0pt}
 \tablehead
 {
 \colhead{No.} &
 \colhead{\#} &
 \colhead{RA} &
 \colhead{Dec} &
 \colhead{\textit{l}} &
 \colhead{\textit{b}} &
 \colhead{Voy} &
 \colhead{UV Bgd} &
 \colhead{Name}
 }
  \startdata
  1 &  793 & 213.9 &  25.4 &  32.0 &  70.5 & 2 &    26.0 & NGC 5548\\
  2 &  820 & 327.4 &  13.9 &  70.7 & -30.0 & 2 &    37.9 & S193 (82\\
  3 &  859 & 359.9 & -43.4 & 331.7 & -71.1 & 2 &    49.3 & EG 165  \\
  4 &  883 &  24.1 & -18.2 & 175.4 & -75.7 & 2 &    49.9 & UV Cet  \\
  5 &  870 &  31.0 & -13.0 & 177.2 & -67.3 & 2 &    54.0 & HD 17925\\
   &    &   &  &     &    &  &    &   \\
396 &  579 & 339.2 &  40.1 &  97.3 & -15.8 & 1 &  3667.9 & 12 Lac B\\
397 &   36 & 134.5 &  58.3 & 158.3 &  39.8 & 1 &  3815.6 & SKY     \\
398 &   37 & 142.6 &  59.5 & 155.2 &  43.5 & 1 &  3928.2 & SKY     \\
399 &  689 & 250.0 & -16.0 &   2.4 &  19.2 & 1 &  4048.4 & OPH-8   \\
400 &  365 & 173.9 &  14.7 & 245.9 &  68.7 & 1 &  4194.1 & Feige 46\\
401 &  201 & 270.4 &  22.4 &  48.5 &  20.1 & 1 &  4230.0 & 97 Her  \\
402 &  449 & 250.4 & -16.3 &   2.4 &  18.7 & 2 &  4375.0 & Drift 1 \\
403 &  744 & 246.2 & -31.1 & 348.1 &  12.1 & 1 &  4445.8 & SKY 3   \\
404 &  503 & 237.0 & -13.5 & 355.5 &  30.3 & 2 &  4710.7 & Drift 3 \\
405 &  531 & 253.0 & -16.2 &   4.1 &  16.8 & 2 &  4730.8 & Drift 6 \\
406 & 1116 & 243.9 & -11.4 &   2.4 &  26.7 & 2 &  4865.5 & SKY 12  \\
407 &  457 & 252.5 & -16.4 &   3.6 &  17.0 & 2 &  5039.5 & Drift 1 \\
408 &  526 & 251.2 & -16.0 &   3.2 &  18.2 & 2 &  5192.4 & Drift 6 \\
409 &  535 & 256.4 & -16.3 &   5.9 &  14.0 & 2 &  5477.9 & Drift 6 \\
410 &  690 & 250.0 & -30.9 & 350.4 &   9.8 & 1 &  5908.8 & OPH-7   \\
411 &  440 & 244.6 & -15.6 & 359.2 &  23.5 & 2 &  6039.3 & Drift 1 \\
412 &  763 & 246.5 & -30.7 & 348.5 &  12.2 & 1 &  6211.4 & SKY 3   \\
413 & 1175 & 435.0 & -11.8 & 211.3 & -29.5 & 2 &  6339.3 & SKY 12  \\
414 &  338 & 215.9 & -62.4 & 313.7 &  -1.7 & 1 &  6509.5 & Pro Cen \\
415 &  524 & 240.4 & -14.1 & 357.5 &  27.5 & 2 &  6653.5 & Drift 5 \\
416 &  518 & 238.5 & -13.8 & 356.4 &  29.2 & 2 &  7286.8 & Drift 5 \\
417 &  443 & 137.6 &  27.8 & 199.1 &  41.9 & 1 &  7328.6 & PG 0909+\\
418 &  438 & 242.5 & -15.3 & 358.1 &  25.2 & 2 &  7452.3 & Drift 1 \\
419 &  521 & 239.4 & -14.0 & 357.0 &  28.3 & 2 &  9223.1 & Drift 5 \\
420 &  999 & 265.8 & -28.9 &   0.1 &  -0.1 & 2 &  9259.8 & G-1     \\
421 &  761 & 251.7 &  -8.2 &  10.2 &  22.4 & 1 &  9542.3 & SKY 5   \\
422 &  984 &  76.2 & -28.1 & 229.8 & -34.1 & 2 &  9980.8 & RS 202 B\\
423 &  714 & 250.5 &  -8.6 &   9.1 &  23.1 & 1 & 10392.9 & SKY 5   \\
424 &  742 & 251.3 &  -8.6 &   9.7 &  22.5 & 1 & 10810.4 & SKY 5   \\
425 &  691 & 250.0 & -26.0 & 354.3 &  13.0 & 1 & 11470.4 & OPH-6   \\
426 &  567 & 112.5 & -22.8 & 237.8 &  -1.9 & 1 & 11635.6 & HD 58978\\
427 &  932 & 193.8 & -62.0 & 303.7 &   0.6 & 2 & 15624.4 & COALSACK\\
428 & 1188 & 189.4 & -64.3 & 301.7 &  -1.7 & 2 & 19249.4 & BKGND 3 \\
429 &  672 & 218.5 &  -1.2 & 348.7 &  51.9 & 1 & 23767.1 & Alp Vir \\
430 &  207 & 237.8 & -20.5 & 350.5 &  24.9 & 1 & 33958.0 & Lam Lib \\
   \enddata
  \end{deluxetable*}
%\clearpage

\acknowledgments

We thank Jay Holberg for his assistance with Voyager over the years.  

\vspace{5mm}
\facilities{GALEX, New Horizons}


\begin{thebibliography}{}
\bibitem[Akshaya et al.(2018)]{2018ApJ...xxx..xxx} Akshaya, M.~S., Murthy, J., Ravichandran, S., Henry, R.~C., \& Overduin, J.\ 2018, \apj, in press 
\bibitem[Ben-Jaffel, L., Holberg, J.(2016)]{2016ApJ...823..161} Ben-Jaffel, L., \& Holberg, J.\ 2016, \apj,  823, 161 
\bibitem[Borthakur(2014)]{2014Science...346..216} Borthakur, S., Heckman, T.~M., et al.\ 2014, Science, 346, 216
\bibitem[Bowyer(1991)]{1991ARAA...29..59} Bowyer, S.\ 1991, ARAA, 29, 59  
\bibitem[Gladstone et al.(2013)]{2013ccfu.book..177G} Gladstone, G.~R., Stern, S.~A., \& Pryor, W.~R.\ 2013, DOI: 10.1007/978-1-4614-6384-9-6 
\bibitem[Gursky et al.(1971)]{1971ApJ...167..L81} Gursky, H., Kellogg, E., Murray, S., Leong, C., Tananbaum, H., \& Giacconi, R.\ 1971, \apjl, 167, L81  
\bibitem[Hamden et al.(2013)]{2013ApJ...779..180} Hamden, E.~T., Schiminovich, D., \& Seibert, M.\ 2013, \apj, 779, 180  
\bibitem[Henry (1973)]{1973ApJ...179..97} Henry, R.~C. \ 1973, \apj, 179, 97  % 32
\bibitem[Henry (1991)]{1991ARAA...29..89} Henry, R.~C.\ 1991, \araa, 29, 89  % 133
\bibitem[Henry(2003)]{} Henry, R.~C.\ 2003, ChJAS, 3, 53H 
\bibitem[Henry et al.(2015)]{2015ApJ...798..14} Henry, R.~C., Murthy, J., Overduin, J., \& Tyler, J.\ 2015, \apj, 798, 14   
\bibitem[Henyey(1941)]{1941ApJ...93..70H} Henyey, L.~G., \&  Greenstein, J.~L.\ 1941, \apj, 93, 70 
\bibitem[Kollmeier et al.(1973ApJL...789..L32)]{} Kollmeier, L. et al.\ 2014, \apjl, 789, L32
\bibitem[Meekins et al.(1971)]{1971Nature...231..107} Meekins, J.~F., Fritz, G., Chubb, T.~A., Friedman, H., \& Henry, R.~C.\ 1971, Nature, 231, 107  
\bibitem[Murthy et al.(1994)]{1994ApJ...428..233} Murthy, J., Henry, R.~C., \& Holberg, J.~B.\ 1994, \apj, 428, 233  
\bibitem[Murthy et al.(1999)]{1999ApJ...522..904} Murthy, J., Hall, D., Earl, M., Henry, R.~C., \& Holberg J.~B.\ 1999, \apj, 522, 904  
\bibitem[Murthy et al.(2012)]{2012ApJS...199..11} Murthy, J., Henry, R.~C., \& Holberg, J.~B.\ 2012, \apjs, 199, 11  
\end{thebibliography}
\end{document}